\documentclass[fleqn,usenatbib]{mnras}

\usepackage{newtxtext,newtxmath}

\usepackage[T1]{fontenc}
\usepackage{orcidlink}

\DeclareRobustCommand{\VAN}[3]{#2}
\let\VANthebibliography\thebibliography
\def\thebibliography{\DeclareRobustCommand{\VAN}[3]{##3}\VANthebibliography}

\usepackage{graphicx}	
\usepackage{amsmath}	

\title[Is the Symbiotic Recurrent Nova T\,CrB Late?]{Is the Symbiotic Recurrent Nova T\,CrB Late? Recent Photometric Evolution and Comparison with Past Pre-Outburst Behaviour}

\author[J. Merc et al.]{\parbox{\textwidth}{Jaroslav Merc$^{1,2}$\thanks{E-mail: jaroslav.merc@mff.cuni.cz (JM)}\orcidlink{0000-0001-6355-2468},
Łukasz Wyrzykowski$^{3,4}$\orcidlink{0000-0002-9658-6151},
Paul G. Beck$^{2,5}$\orcidlink{0000-0003-4745-2242},
Przemysław J. Mikołajczyk$^{3,4,6}$\orcidlink{0000-0001-8916-8050},
Krzysztof Kotysz$^{3,6}$\orcidlink{0000-0003-4960-7463},
Paweł Zieliński$^{7}$\orcidlink{0000-0001-6434-9429},
Staszek Zola$^{8}$\orcidlink{0000-0003-3609-382X},
Sebastian Kurowski$^{8}$\orcidlink{0000-0002-1557-0343},
Waldemar Ogłoza$^{9}$\orcidlink{0000-0002-6293-9940},
Marek Drozdz$^{9}$,
Charles Galdies$^{10,11}$\orcidlink{0000-0002-8908-0785},
Franz-Josef Hambsch$^{12,13}$\orcidlink{0000-0003-0125-8700},
Stephen M. Brincat$^{14}$\orcidlink{0000-0002-9205-5329},
Barbara Joachimczyk$^{7}$\orcidlink{0009-0005-1710-6754},
Mateusz Bronikowski$^{15}$\orcidlink{0000-0002-1537-6911},
Jure Japelj$^{15}$,
Matej Mihelcic$^{16}$,
Josep Manel Carrasco$^{17,18,19}$\orcidlink{0000-0002-3029-5853},
Umut Burgaz$^{20}$\orcidlink{0000-0003-0126-3999},
Agnieszka Gurgul$^{7}$\orcidlink{0000-0002-9441-0195},
Karolina Bąkowska$^{7}$\orcidlink{0000-0003-1034-1557},
Piotr Hofbauer$^{7}$,
Krzysztof Szyszka$^{7}$\orcidlink{0009-0007-5650-2818},
Jan Golonka$^{7}$\orcidlink{0009-0001-7100-5218},
Jan Kåre Trandem Qvam$^{21}$\orcidlink{0009-0008-6794-5099},
Justas Zdanavičius$^{22}$\orcidlink{0009-0000-9910-1124},
Erika Pakštienė$^{22}$\orcidlink{0000-0002-3326-2918},
Marius Maskoliūnas$^{22}$\orcidlink{0000-0003-3432-2393},
Vytautas Čepas$^{22}$,
Uliana Pylypenko$^{3}$\orcidlink{0009-0002-7560-1903},
Dawid Moździerski$^{6}$\orcidlink{0000-0002-3861-9031},
Franky Dubois$^{23,12}$,
Siegfried Vanaverbeke$^{23,12}$\orcidlink{0000-0003-0231-2676},
Justyna M. Olszewska$^{24}$\orcidlink{0000-0003-3579-4253},
Alexios Liakos$^{25}$\orcidlink{0000-0002-0490-1469},
Milan Stojanović$^{26}$\orcidlink{0000-0002-4105-7113},
Goran Damljanović$^{26}$\orcidlink{0000-0002-6710-6868},
Adam Popowicz$^{27}$\orcidlink{0000-0003-3184-5228},
Mateusz Marzec$^{8}$,
Magdalena Badura$^{8}$,
Bartosz Gil$^{8}$,
Alicja Pucek$^{8}$\orcidlink{0009-0009-2238-6913},
Aleksandra Kowalska$^{8}$,
Mateusz Szklarz$^{8}$,
Teimuraz Kvernadze$^{28}$\orcidlink{0000-0001-9947-4983},
Andrea Reguitti$^{29,30}$\orcidlink{0000-0003-4254-2724},
Supachai Awiphan$^{31}$\orcidlink{0000-0003-3251-3583},
Michel Dennefeld$^{32}$\orcidlink{0000-0002-8197-5410},
Kosmas Gazeas$^{33}$\orcidlink{0000-0002-8855-3923}
\newline
\emph{\normalsize Affiliations are listed at the end of the paper}
}}

\date{Accepted 2025 May 06. Received 2025 May 02; in original form 2025 April 29}

\pubyear{\the\year{}}

\begin{document}
\label{firstpage}
\pagerange{\pageref{firstpage}--\pageref{lastpage}}
\maketitle

\begin{abstract}
T\,CrB is a symbiotic recurrent nova that last erupted in 1946. Given its recurrence timescale of approximately 80 years, the next outburst is eagerly anticipated by the astronomical community. In this work, we analyse the optical light curves of T\,CrB, comparing recent photometric evolution with historical data to evaluate potential predictive indicators of nova eruptions. Although the "super-active" phases preceding both the 1946 and anticipated eruptions are strikingly similar, the subsequent photometric behaviour differs. We find that the decline in brightness observed in 2023, interpreted by some as a "pre-eruption dip", deviates from the deep minimum recorded prior to the 1946 event and does not reliably predict the eruption timing. Recent photometric and spectroscopic observations indicate that the system is returning to a high-accretion state. Given this, an eruption may be imminent, even without distinct precursors. While the next eruption of T\,CrB will be a major scientific event, its expected peak brightness of $V \sim 2$~mag highlights the importance of setting realistic public expectations for what will be a visually modest, yet astrophysically very significant, celestial event.

\end{abstract}




\section{Introduction}
Nova eruptions are driven by thermonuclear runaways on the surface of white dwarfs accreting material from a binary companion. In principle, all novae are recurrent, as the eruptions can repeat once sufficient mass has been accreted. However, recurrence timescales can span many centuries or longer, and only those systems that have exhibited multiple eruptions within observational timescales, typically decades to a century, are classified as recurrent novae (RNe). For comprehensive reviews on novae and RNe, see, e.g., \citet{2021ARA&A..59..391C} and \citet{2021gacv.workE..44D}.

T~Coronae~Borealis (T\,CrB) is one of eleven known RNe in the Milky Way \citep[][]{2021gacv.workE..44D,2022MNRAS.517.6150S,2022MNRAS.517.3864S,2024MNRAS.529..224S} and one of only four hosting a red giant donor (T\,CrB, RS\,Oph, V745\,Sco, and V3890\,Sgr), making it a member of the symbiotic subclass of RNe. Among RNe, it is the brightest in quiescence \citep[$V\sim9.8$ mag;][]{2023MNRAS.524.3146S}. The binary system consists of a massive white dwarf that accretes material from an M4III-type red giant companion, in a circular orbit with a period of 227.6 days \citep{2025A&A...694A..85P,2025ApJ...983...76H}. The mass of the white dwarf has been estimated to be close to the Chandrasekhar limit \citep[$M_\mathrm{WD} \approx 1.37\,M_\odot$;][]{2025ApJ...983...76H}, while the red giant, with a mass of $\sim 0.69\,M_\odot$ and a radius of 65\,$R_\odot$ \citep[][]{2025ApJ...983...76H} fills its Roche lobe, as evidenced by prominent ellipsoidal variability (with an amplitude of $\sim$0.3 mag in $V$). The system lies at a distance of approximately 890~pc \citep[][]{2021AJ....161..147B,2023A&A...674A...1G}.

T\,CrB has experienced two confirmed, very fast nova eruptions, in 1866 and 1946, both reaching a visual magnitude of approximately $V \sim 2$~mag \citep[][]{1946PASP...58..153P, 2023MNRAS.524.3146S}. The sharp rise (about 8 mag from the quiescent level to peak brightness) occurred on a timescale of just a few hours, followed by a rapid decline in brightness, with the system fading by 2, 3, and 6 magnitudes within 3, 5, and 12 days, respectively \citep[][]{2023MNRAS.524.3146S}. Historical records also suggest a highly probable earlier outburst in 1787 and a more uncertain event recorded in 1217 \citep{2023JHA....54..436S}. The $\sim$80-year interval between the confirmed eruptions provides a tentative recurrence timescale, indicating that a new outburst may occur very soon.

The system is currently being monitored by numerous observational campaigns, driven by similarities between its recent photometric and spectroscopic behaviour and that observed prior to the 1946 eruption (mainly so called "super-active" phase and the "pre-eruption dip"). These parallels have led several authors to conclude that another nova outburst is imminent, with predicted eruption dates ranging between 2023 and 2026 \citep[e.g.][]{2016NewA...47....7M, 2020ApJ...902L..14L, 2023AstL...49..501M, 2023MNRAS.524.3146S, 2024MNRAS.532.1421T}.



As a result, a global observing effort is currently underway, involving both professional and amateur astronomers, with the aim of capturing the next eruption in unprecedented detail across all accessible wavelengths and with virtually every available instrument. This will yield data with a level of precision, cadence, and wavelength coverage that has never been achieved for previous eruptions of T\,CrB, nor for any other object of this class, opening a unique window for new discoveries and for explaining the peculiarities of T\,CrB. Among other things, T\,CrB is predicted to become the brightest classical or recurrent nova ever observed in X-rays, with two expected X-ray episodes \citep[][]{2025ApJ...982...89S}, including an early X-ray flash, so far seen only in the classical nova YZ\,Ret \citep[][]{2022Natur.605..248K}. Due to strong shocks, it is also expected to be a prominent source of very high-energy $\gamma$-rays, similar to RS\,Oph \citep[][]{2022Sci...376...77H, 2022NatAs...6..689A}.

Infrared spectroscopy, including integral field unit spectrographs, can be used to study nucleosynthesis and the structure of the ejecta \citep[see][]{2025ApJ...982...89S}, while infrared interferometry will allow spatially resolved studies of the ejecta as it expands \citep[e.g.,][]{2025CoSka..55c..67O}. High-resolution optical and infrared imaging may also enable the detection of light echoes from the recently discovered nova super-remnant surrounding T\,CrB \citep[][]{2024ApJ...977L..48S}. While fluorescent light echoes are unlikely to be detectable according to the authors, dust-scattered continuum echoes might be observed, probing the distribution and properties of the circumstellar dust.

It will also be interesting to investigate the impact of the eruption on the orbital period of the system. \citet{2023MNRAS.524.3146S} reported a significant period change after the 1946 eruption based on photometric data. Now, for the first time, we will have the opportunity to compare precise spectroscopic orbital solutions before and after an eruption. Finally, the anticipated brightness and accessibility of the system will also make it a unique opportunity for public engagement in transient astronomy.

In this work, we present and analyse new photometric observations of T\,CrB obtained with the global network of telescopes coordinated using BHTOM.space\footnote{\url{https://bhtom.space}} tool, complemented by archival data from the database of the American Association of Variable Star Observers\footnote{\url{https://www.aavso.org}} (AAVSO) and the literature. Our aim is to characterise the current behaviour of the system in the context of its historical evolution, with particular attention to the super-active phase and reported pre-outburst signatures. We show that if the evolution during the super-active phase is indeed causally linked to the nova eruption and follows a consistent pattern between outbursts, the eruption should have already occurred. Conversely, the decline in brightness observed around 2023, interpreted by some authors as a pre-eruption dip, does not exhibit the same characteristics as the evolution preceding the 1946 eruption.

\section{Observations}

\subsection{BHTOM}
\label{sec:bhtom} 

BHTOM (Black Hole Target and Observation Manager) is an open-access, community-driven, web-based system for managing and coordinating time-domain astrophysical observations, built upon the Las Cumbres Observatory's TOM Toolkit \citep{2018SPIE10707E..11S}. Designed to be flexible and extensible, BHTOM enables researchers to efficiently track, prioritize, and follow up on a wide variety of transient and variable phenomena.

It supports a diverse range of targets, including supernovae, tidal disruption events, X-ray binaries, novae, gravitational microlensing events, variable stars, and quasars. BHTOM integrates tools for scheduling observations, ingesting new alerts, and visualizing and analysing light curves. It can automatically process images from more than 130 telescopes worldwide, ranging from 0.2-m to 2.5-m in diameter, by performing high-quality PSF photometry and standardizing the results to \textit{Gaia} Synthetic Photometry. For further details and examples of use, see e.g., \cite{2019CoSka..49..125Z,2020A&A...633A..98W,2020A&A...644A..49M,2022A&A...657A..18R,2023MNRAS.524.3344N,2024AcA....74...77M}.

\begin{table}
	\centering
	\caption{Total number of observations in each photometric band and the corresponding range of covered Julian Dates (JD $-2\,400\,000$). }
	\label{tab:observations}
	\begin{tabular}{rrrrr} 
		\hline
		Filter & BHTOM & Range & AAVSO & Range\\
		\hline
		\textit{B} & 2\,524 & 59\,927--60\,762 & 119\,289 & 53\,076--60\,782\\
		\textit{V} & 5\,916 & 59\,927--60\,766 & 121\,147 & 41\,833--60\,782\\
		\textit{R} & 3\,004 & 59\,927--60\,766 & 19\,022 & 53\,823--60\,781\\ 
		\textit{I} & 3\,629 & 59\,927--60\,766 & 14\,729 & 53\,076--60\,780\\
		\textit{Vis.} & - & - & 139\,012 & 2\,744--60\,782\\
		\hline
	\end{tabular}
\end{table}

BHTOM is a free and open platform that serves the global astronomical community by optimizing follow-up efforts and maximizing the scientific return from both new observations and historical data extracted from numerous public surveys. Thanks to its simplicity and automation in data handling and processing, BHTOM is a valuable tool for both professional astronomers and amateur observers alike.

T\,CrB has been monitored by the BHTOM network of telescopes since December 2022. In this work, we use data collected through the end of March 2025 (see Table \ref{tab:observations}). Table~\ref{tab:bhtom} lists all telescopes that contributed to the BHTOM dataset. Although the data were collected using a variety of filters, we only utilize observations standardized to the Johnson–Kron-Cousins \textit{BVRI} system in this study.

The BHTOM data were cleaned by removing measurements affected by saturation. Outliers were filtered by modeling the light curves in each band with a sinusoid corresponding to the period of ellipsoidal variability (i.e., half the orbital period of the system), with the time of minimum set to MJD 59096.6031. Data points deviating by more than 0.25 mag from the fitted curve were excluded. This filtering step rejected approximately 10\% of the measurements in each band.

\subsection{Other sources}
Our photometry is complemented by \textit{B}, \textit{V}, \textit{R}, \textit{I}, and visual observations from the database of AAVSO \citep[][]{Kloppenborg2025}. In total, the light curves comprise 413\,199 individual observations (see Table~\ref{tab:observations}), with the visual observations available since 1866.

Historical data are taken from the comprehensive compilation by \citet{2023MNRAS.524.3146S}, which includes literature \textit{B} and \textit{V} observations, additional visual data, and photographic measurements calibrated to the \textit{B} band from archival plates from the Harvard College Observatory, Bamberg Observatory, and Sonneberg Observatory. Combined with the AAVSO data and our own observations, these data trace the photometric evolution of T\,CrB from its outburst in 1866 to the present.

\begin{table*}
\caption{List of the BHTOM telescopes contributing to this work.}
\begin{tabular}{lrl}
\hline
 Telescope                     &   Diameter [m] & Location                                                           \\
\hline
 ASV Telescope Milanković      &     1.40   & Astronomical Station Vidojevica, Astronomical Observatory, Belgrade, Serbia   \\
 60-cm ASV telescope           &     0.60   & Astronomical Station Vidojevica, Astronomical Observatory, Belgrade, Serbia             \\
 ASV Telescope Meade           &     0.40   & Astronomical Station Vidojevica, Astronomical Observatory, Belgrade, Serbia   \\
 Sky-watcher quattro  &     0.25  & Adonis Observatory, Belgium                                        \\
 ASA DM160                     &     0.60   & Adiyaman Observatory, Turkey                                       \\
 Schmidt 67/92 Telescope       &     0.91  & Padova Astronomical Observatory, Italy                             \\
 Białków Large Telescope &     0.60   & Astronomical Institute, University of Wroclaw, Poland              \\
 1.23-m telescope on Calar Alto &     1.23  & Calar Alto Astronomical Observatory, Spain                         \\
 Meade SSC-10                  &     0.25  & Flarestar Observatory, Malta                                       \\
 GeoNAO SCTelescope     &     0.36 & Georgian National Astrophysical Observatory, Georgia               \\
 GoChile GoT1                  &     0.40   & El Sauce Observatory, Chile                                        \\
 Horten 0.68-m                 &     0.68 & Horten Local Observatory, Norway                                   \\
 Kryoneri telescope    &     1.20   & Kryoneri Observatory, National Observatory of Athens, Greece \\
 Las Cumbres Observatory HO40                    &     0.40   & Haleakala High Altitude Observatory, United States                 \\
 Las Cumbres Observatory MCD40                   &     0.40   & McDonald Observatory, United States                                \\
 Las Cumbres Observatory TO40                    &     0.40   & Teide Observatory, Spain                                           \\
 35-cm Maksutov                &     0.35  & Moletai Observatory, Lithuania                                     \\
 OAUJ-CDK500                   &     0.50   & Obserwatorium Astronomiczne Uniwersytetu Jagiellońskiego, Poland   \\
 TSC90                         &     0.90   & Piwnice Observatory, Institute of Astronomy of the Nicolaus Copernicus University in Toruń, Poland                                                    \\
 CDK700                        &     0.70   & Planetarium Slaskie, Poland                                        \\
 R-COP Celestron C14           &     0.35  & Perth Observatory, Australia                                       \\
 ROAD ODK 40-cm f6.8                &     0.40   & Remote Observatory Atacama Desert, Chile                           \\
 RRRT                          &     0.60   & Fan Mountains Observatory, United States                           \\
 SARA-KP                       &     0.91 & Kitt Peak, United States                                           \\
 SOA RC16                      &     0.41 & Szkolne Obserwatorium Astronomiczne, Bolęcina, Poland                                                        \\
 0.6-m Cassegrain Zeiss        &     0.60   & Mt. Suhora Observatory, Poland                                     \\
 SUTO-Otivar                   &     0.30   & Silesian University of Technology Observatories, Spain             \\
 Telescope Joan Oró  &     0.80   & The Montsec Astronomical Observatory, Spain                        \\
 20-cm SCT Telescope           &     0.20   & Znith Astronomy Observatory, Malta                                 \\
\hline
\end{tabular}
\label{tab:bhtom}
\end{table*}

\section{Recent evolution}\label{sec:recent}

\begin{figure*}
	\includegraphics[width=0.85\textwidth]{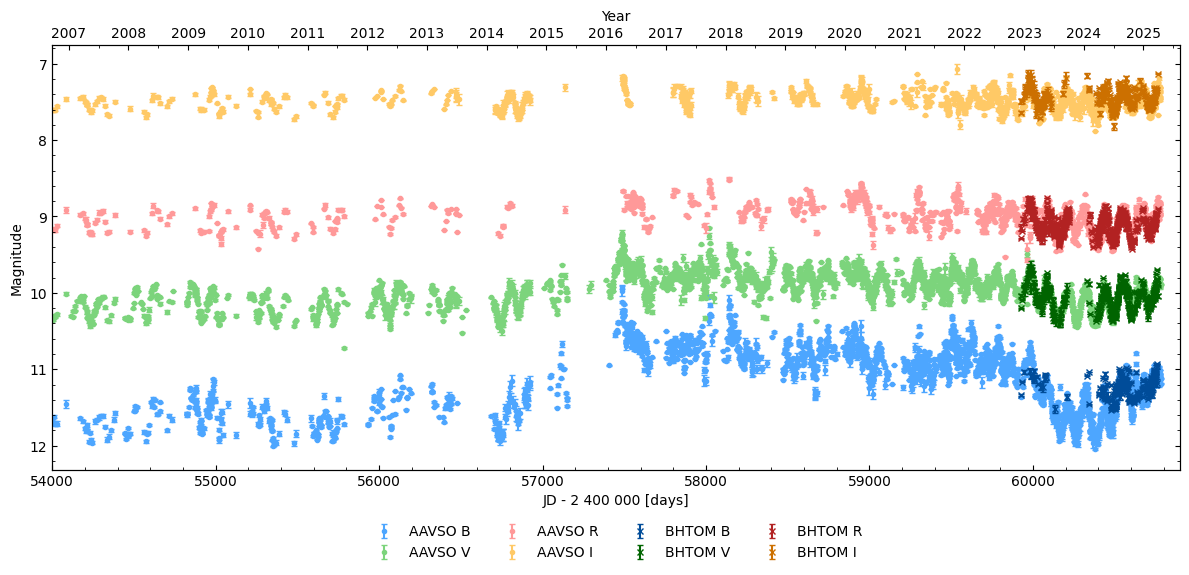}
    \caption{Recent photometric evolution of T\,CrB. Light curves in the \textit{B, V, R}, and \textit{I} bands obtained from AAVSO and BHTOM are shown. The data have been averaged over 1-day intervals \citep[to reduce the scatter due to flickering, that is always present in the light curves; see, e.g., ][and references therein]{2023ApJ...953L...7I,2024A&A...683A..84M} following the removal of outliers.}
    \label{fig:recent}
\end{figure*}

We are conducting regular photometric monitoring of T\,CrB, with data processed through the BHTOM system (see Section \ref{sec:bhtom}). Observations in the \textit{B}, \textit{V}, \textit{R}, and \textit{I} bands are shown in Fig.~\ref{fig:recent}.

Figure~\ref{fig:recent} presents the photometric evolution of T\,CrB from late 2006 to 2025. At longer wavelengths (\textit{R} and, particularly, \textit{I}), the light curves are dominated by ellipsoidal variability throughout the entire interval. This variability arises from the tidal distortion of the Roche-lobe-filling giant and results in a modulation with a period of half the orbital period. The super-active accretion phase, that began in late 2014 and ended in 2023 \citep[][]{2016NewA...47....7M, 2023RNAAS...7..145M}, is clearly visible at shorter wavelengths, most notably in the \textit{B} band, where the system appears more than 1 mag brighter on average compared to the preceding years. Several individual brightness maxima are evident and are independently confirmed by other studies \citep[e.g.,][]{2023RNAAS...7..145M,2024ATel16404....1M}. From 2022 onwards, ellipsoidal variations once again become clearly detectable in the \textit{B} light curve. After returning to quiescent brightness levels in 2023, the system began to brighten again in 2024, though it has not yet reached the levels characteristic of the previous super-active phase.

The increased brightness during the super-active stage is also present in the \textit{V} band, albeit with smaller amplitude, and the ellipsoidal modulation remains detectable throughout most of the interval. This is consistent with the larger contribution of the red giant to the total flux in \textit{V}, where it is $\sim$1.6 mag brighter than in the \textit{B} band \citep[][]{2023ApJ...953L...7I}. Some individual flares are also discernible in the \textit{V} light curve, most notably the event observed in 2016. After the end of the super-active phase, the system again reached quiescent brightness in \textit{V}, followed by a slight rise in 2024.

A comparison between the \textit{B} and \textit{V} light curves confirms that enhanced activity phases are more prominent at shorter optical wavelengths. Interestingly, in addition to the super-active phase, the \textit{B}-band data reveal two earlier episodes of moderately increased brightness around 2009 and 2012. While the system did not by far reach the same magnitudes as during the super-active phase, these intervals stand out clearly above the surrounding quiescent levels and are not evident in the other filters.

\section{Discussion}

\begin{figure*}
\centering
	\includegraphics[width=0.85\textwidth]{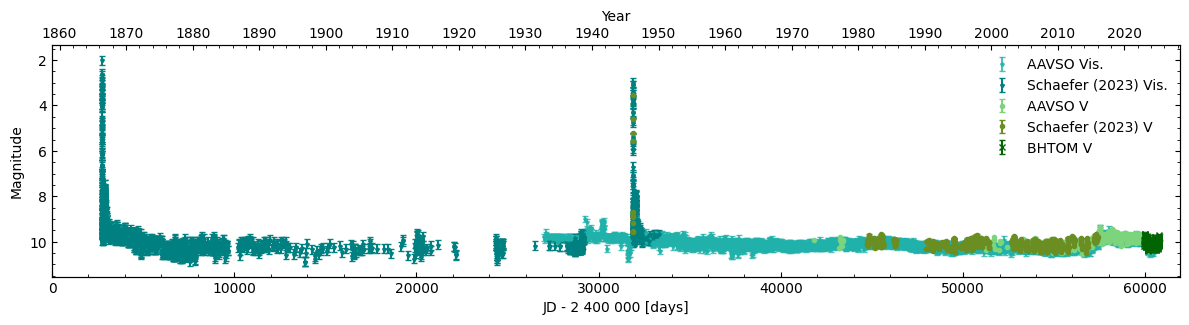}
    \includegraphics[width=0.85\textwidth]{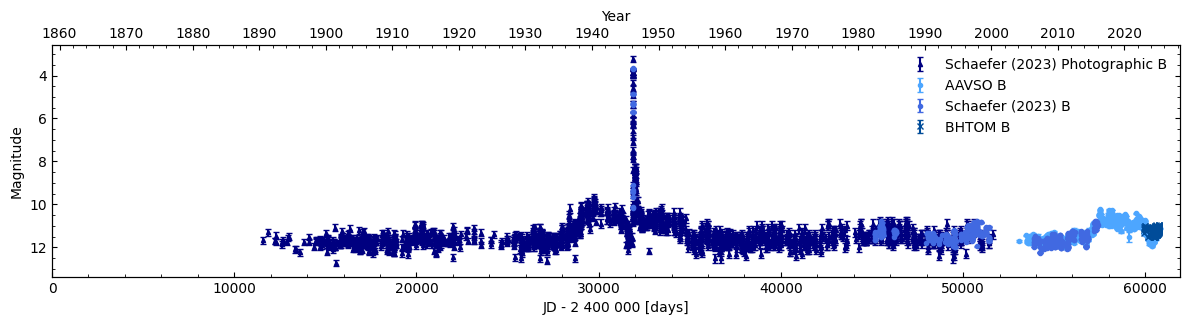}
    \caption{Long-term photometry of T\,CrB in the \textit{V} band (upper panel) and \textit{B} band (lower panel), covering the period from the 1866 outburst to 2025. AAVSO observations have been averaged over 10-day intervals. All data from \citet{2023MNRAS.524.3146S} are shown, with the exception of AAVSO data collected by the author, which are taken directly from the AAVSO database.}
    \label{fig:long_term}
\end{figure*}
\subsection{The super-active phase as a predictor of the nova outburst}
The long-term light curves of T\,CrB in the \textit{V} and \textit{B} bands, covering the period from 1866 to the present, are shown in Fig.~\ref{fig:long_term}. Both nova outbursts are clearly visible in the \textit{V}-band data. The \textit{B}-band observations span a large portion of the quiescent period preceding the 1946 eruption, the eruption itself, and continue almost uninterrupted to the present day. The photometric evolution during both nova events was strikingly similar, with each outburst followed, after approximately 80 days, by a rebrightening of smaller amplitude and lasting about 100 days \citep[see e.g. figures 2 and 3 in][]{2023MNRAS.524.3146S}. The secondary maximum has been interpreted either as the result of an irradiated, tilted disk \citep[][]{1999ApJ...517L..47H} or, more recently, as irradiation of the red giant donor by the cooling white dwarf \citep[][]{2023RNAAS...7..251M}.

\begin{figure}
    \centering
	\includegraphics[width=0.72\columnwidth]{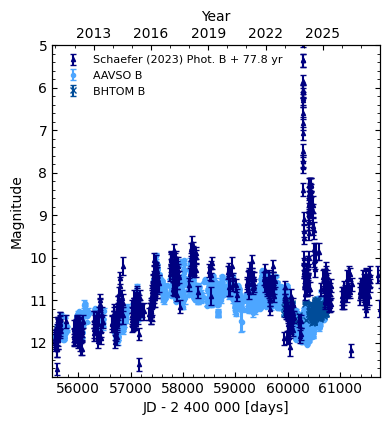}
    \caption{Comparison of the recent behaviour of T\,CrB in the \textit{B} band (light and dark blue)  with its pre-outburst evolution prior to the 1946 outburst (navy blue). The historical photographic data are taken from \citet{2023MNRAS.524.3146S} and have been shifted by +77.8 yr (28\,430 days) to align the super-active phases observed before the 1946 outburst and in the present epoch. Recent data are taken from AAVSO and BHTOM.}
    \label{fig:zoom_in_B}
\end{figure}

The super-active phase observed $\sim$10 years before and $\sim$9 years after the 1946 eruption \citep[][]{2023MNRAS.524.3146S} is clearly visible in the \textit{B} light curve. This active stage was interrupted only by the pre-eruption dip, the eruption itself, and the subsequent secondary maximum. A comparable rise in brightness has also been observed in recent years, beginning in 2014 and ending in 2023, and was interpreted as a possible precursor to the next eruption. By aligning the recent and historical light curves, some authors have attempted to predict the timing of the anticipated outburst.

A detailed comparison of the current and historical \textit{B} light curves during these active phases is shown in Fig.~\ref{fig:zoom_in_B}, where the historical data have been shifted by 77.8 years (28\,430 days) to align the super-active phases. The similarity in photometric evolution between the two epochs is apparent. However, if this phase is causally linked to nova eruptions and follows a similar timescale, it would imply that the eruption should have already occurred around 2024. This suggests that although the photometric behaviour observed in the decades surrounding the 1946 eruption and the present day is qualitatively similar, the detailed evolution differs to some extent. Consequently, predicting the exact time of the next eruption based on such comparisons remains challenging and demonstrates the incomplete understanding of the physical processes behind the super-active phase.

In a recent study, \citet{2023ApJ...953L...7I} proposed that the active phases of T\,CrB resemble the outbursts and super-outbursts seen in SU~UMa-type dwarf novae. According to their analysis, larger active phases recur approximately every $\sim$5000 days, while smaller ones occur on a $\sim$1000-day timescale \citep[][]{2016MNRAS.462.2695I,2023ApJ...953L...7I}. These stages of increased brightness are visible in the \textit{B} light curve in Fig.~\ref{fig:long_term}, with the ones around 1980, 1997, and the recent super-active phase reported by \citet{2023ApJ...953L...7I}. Another brightening episode around 1966 also fits the longer recurrence period. Examples of smaller active phases are visible in Fig.~\ref{fig:recent}, as discussed in Sec.~\ref{sec:recent}.

\citet{2020ApJ...902L..14L} argued that during these high-accretion phases, T\,CrB accretes a significant fraction of the mass required to ignite a thermonuclear runaway, whereas the average accretion rate in quiescence is too low to support nova outbursts on an $\sim$80-year timescale. Based on \textit{UBV} photometry, \citet{2023A&A...680L..18Z} estimated that approximately 30\% of the necessary mass was accumulated during the recent super-active phase. A similar conclusion follows from the accretion rate derived from optical and X-ray data by \citet{2024MNRAS.532.1421T}. Given that the recent super-active phase was the most luminous since the one surrounding the 1946 eruption, it is reasonable to assume that a large portion of the ignition mass was accreted during this time.

Taken together, it is perhaps no surprise that T\,CrB can reach the ignition threshold at some point during a prolonged high-accretion stage (lasting $\sim$19 years in the previous cycle), even if these two phenomena are not directly causally linked. However, the precise timing of the nova eruption also depends on the amount of mass accreted outside the major active phases, both in the post-1946 period and more recently, making it difficult to determine the exact eruption date.

\subsection{On the reported pre-eruption dip}
The 1946 nova outburst was immediately (starting $\sim$1 yr before the eruption) preceded by a sharp decline in brightness, referred to as the "pre-eruption dip" \citep{2023MNRAS.524.3146S}. This phenomenon is particularly puzzling because, in the \textit{V} band, where the dip is most pronounced, the observed brightness dropped to more than 1.5 mag below the typical quiescent level of the red giant donor itself. To explain this, \citet{2023MNRAS.524.3146S} proposed circumstellar dust obscuration as the likely cause. However, such an interpretation would imply an even stronger dimming in the bluer bands, which was not observed. The explanation thus remains an open question.

\begin{figure}
    \centering
	\includegraphics[width=0.72\columnwidth]{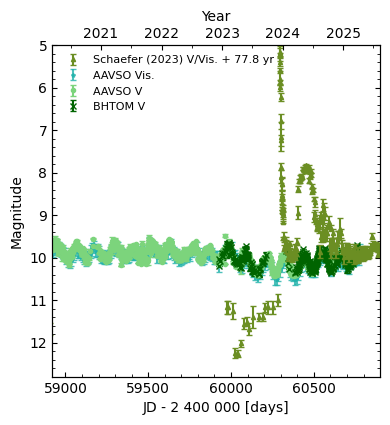}
    \caption{Recent evolution of T\,CrB in the \textit{V} band (light and dark green) compared with its behaviour prior to the 1946 outburst (olive green). The same temporal shift of +77.8 yr (28\,430 days) as in Fig.~\ref{fig:zoom_in_B} has been applied to align the ingress and decline of the super-active phases observed in the \textit{B} band. Recent data are taken from AAVSO and BHTOM, while the historical data are from \citet{2023MNRAS.524.3146S} and represent a compilation of all available observations in the \textit{V} and visual bands from the literature and AAVSO.}
    \label{fig:zoom_in_V}
\end{figure}

The same author argued that if a similar dip were to occur prior to a future outburst, it could serve as an excellent predictor of the eruption, with the nova expected to follow after about a year. In March--April 2023, to the excitement of the astronomical community, T\,CrB began to decline in brightness. This fading was interpreted by some as a new pre-eruption dip \citep[][]{2023ATel16107....1S,2023BAAVC.196....8T}, and based on this assumption, \citet{2023ATel16107....1S} predicted the eruption to occur around 2024.4$\pm$0.3. No outburst was observed, though.

This decline, marking the apparent end of the recent super-active phase \citep[][]{2023RNAAS...7..145M}, is clearly visible in both \textit{B} and \textit{V} light curves of T\,CrB. To compare the current evolution with historical data, we again shifted the archival light curve by 77.8 years (28\,430 days). As shown in Fig.~\ref{fig:zoom_in_B}, the decline in the \textit{B} band aligns remarkably well with the one observed before the 1946 eruption. However, the behavior in the \textit{V} band is notably different (Fig.~\ref{fig:zoom_in_V}). This time, no deep decline is observed; instead, T\,CrB returned to its typical quiescent brightness level before beginning to brighten again (see Sec.~\ref{sec:recent}).

The comparison between the recent and historical light curves suggests three possible interpretations. First, there may be no direct causal link between the nova outburst and the pre-eruption dip. In that case, similar behaviour may not necessarily precede future eruptions and the absence of a dip would not preclude an imminent outburst. Second, the observed minimum in 2023 might represent a pre-eruption dip, but its evolution significantly differs from the deep and prolonged dip observed before the 1946 outburst. This suggests that such a feature, even if present, does not provide a reliable predictor of eruption timing. Third, it is possible, although probably unlikely, that the “true” pre-eruption dip, regardless of its physical origin, has not yet occurred. If this feature does in fact act as a consistent precursor to eruption, then the nova outburst is unlikely to occur within the next year. It is worth noting that similar behaviour has not been confirmed in any other nova to date, with only a tentative hint of a very mild decline reported by \citet{2023arXiv230804104Z} before the outburst of RS\,Oph.

\subsection{When will T\,CrB erupt?}
It appears challenging to precisely predict the timing of the next eruption based on comparisons between the current behaviour of T\,CrB and its historical pre-outburst evolution. If one considers only the timing of the super-active phases, the eruption should have already occurred. The same conclusion follows if the recent decline in brightness is interpreted as the pre-eruption dip.

The interval between the peaks of the 1866 and 1946 outbursts was 29\,125 days, or 79.7 years \citep[][]{2023MNRAS.524.3146S}, which would place the next eruption in November 2025 if the exactly same recurrence time is assumed. Alternatively, if the 18th-century eruption occurred around December 20, 1787, as proposed by \citet[][]{2023JHA....54..436S}, then the time elapsed until the 1866 outburst was approximately 28\,633 days (78.4 years), which would predict the next eruption in July 2024. Taking the average of these two intervals suggests an expected eruption around March 2025. Furthermore, \citet[][]{2023JHA....54..436S} calculated that if the transient observed in 1217 also corresponds to a T\,CrB outburst, the average recurrence time from 1217 to 1946 would be about 80.9 years. This would place the expected nova eruption near the boundary of 2026 and 2027.

However, the recurrence intervals discussed above clearly demonstrate that T\,CrB (and recurrent novae in general) do not follow a strictly periodic pattern. The recent rise in brightness suggests that T\,CrB is once again entering a phase of elevated accretion. This interpretation is supported by spectroscopic observations, which show the reappearance and strengthening of emission lines (e.g., Balmer lines of \ion{H}{i}, \ion{He}{ii}) \citep[e.g.,][]{2024ATel16912....1T,2025ATel17030....1S,2025ATel17075....1H,2025ATel17110....1M}. These features had previously vanished or become significantly weaker following the conclusion of the super-active phase in 2023 \citep[][]{2023RNAAS...7..145M,2025BlgAJ..42...29S}.

Given that the system is expected to have already accreted most of the mass required for ignition, it is plausible that T\,CrB will soon reach the threshold, maybe without any clear early warning.

The timing of the upcoming eruption may also influence the timing of the subsequent nova event in the early years of the next century. If the post-outburst high-accretion phase is shorter than the one that followed the 1946 eruption (assuming a similar total duration of around 19 years, but with the anticipated eruption occurring later within this phase) then the next eruption of T\,CrB could occur at the end of, or even after, a comparable future high-accretion stage, should such a stage recur again in about 80 years.

\section{Conclusions}
The comparison of the recent photometric evolution of T\,CrB with its historical pre-outburst behaviour, particularly that preceding the 1946 eruption, reveals that although the super-active phases were remarkably similar, the subsequent evolution diverged. If the recent decrease in brightness represents a pre-eruption dip, its characteristics differ significantly from those observed prior to the 1946 outburst. Neither the super-active phase nor the recent fading event have, so far, reliably predicted the onset of the eruption. At the same time, the system now appears to be entering a new phase of elevated accretion. Given that it is likely already near the critical ignition threshold, this may indicate that the eruption is imminent, though it could arrive without distinctive or predictable photometric signatures.

In considering all the predictions and analyses, it is also important to acknowledge an aspect that extends beyond the scientific domain. The next eruption of T\,CrB is not only a highly anticipated event within the astronomical community, but also one that has captured the attention of the wider public and media. This offers a unique opportunity to engage the public in the field of time-domain and transient astronomy. 

However, it is crucial to emphasize that predicting the exact timing of such an event remains inherently uncertain. While there have been attempts (particularly in the non-refereed literature) to assign a specific date to the eruption, these should be approached with caution. Headlines forecasting "an extraordinary celestial performance visible next Wednesday" may not reflect the scientific reality.

Moreover, while the eruption of T\,CrB will undoubtedly be a rare and scientifically significant event, monitored across the electromagnetic spectrum and likely among the brightest transients seen in decades, public expectations should be tempered with realistic context. The nova is expected to reach a peak brightness of approximately $V \sim 2$~mag, comparable to Polaris, but also to less widely known stars such as Alphard, Hamal, or Diphda. At the time of maximum light, T\,CrB will still be fainter than roughly 50 other stars in the sky, and the eruption will not feature any dramatic visual phenomena. Referring to it as a celestial “firework” is therefore misleading. Ensuring that the public receives accurate and appropriately scaled information is essential for fostering a meaningful and informed appreciation of this remarkable astronomical event.

\section*{Acknowledgements}
We thank the referee, Michael Shara, for a careful reading of the manuscript and for the constructive suggestions that helped improve it. We acknowledge with thanks the variable star observations from the AAVSO International Database contributed by observers worldwide and used in this research.

The research of J.M. was supported by the Czech Science Foundation (GACR) project no. 24-10608O. J.M. and P.G.B. acknowledge support by the Spanish Ministry of Science and Innovation with the grant no. PID2023-146453NB-100 (PLAtoSOnG). P.G.B. acknowledges support by the Spanish Ministry of Science and Innovation with the \textit{Ram{\'o}n\,y\,Cajal} fellowship number RYC-2021-033137-I and the number MRR4032204.

BHTOM.space is based on the open-source TOM Toolkit by LCO and has been developed with funding from the OPTICON-RadioNet Pilot (ORP) of the European Union's Horizon 2020 research and innovation programme under grant agreement No 101004719 (2021-2025). This project has received funding from the European Union's Horizon Europe Research and Innovation programme ACME under grant agreement No 101131928 (2024-2028). {\L}.W. acknowledges support from the Polish National Science Centre DAINA grant No 2024/52/L/ST9/00210. 

M.B. acknowledges the financial support from the Slovenian Research Agency (research core funding P1-0031, infrastructure program I0-0033). GoChile is a project funded by the University of Nova Gorica and the Astronomical Magazine Spika. 

J.M.C. was (partially) supported by the Spanish MICIN/AEI/10.13039/501100011033 and by "ERDF A way of making Europe" by the “European Union” through grant PID2021-122842OB-C21, and the Institute of Cosmos Sciences, University of Barcelona (ICCUB, Unidad de Excelencia ’Mar\'{\i}a de Maeztu’) through grant CEX2019-000918-M. The Joan Or\'o Telescope (TJO) of the Montsec Observatory (OdM) is owned by the Catalan Government and operated by the Institute for Space Studies of Catalonia (IEEC).

Some of the observations have been obtained with the 90cm Schmidt-Cassegrain Telescope (TSC90) in Piwnice Observatory, Institute of Astronomy of the Nicolaus Copernicus University in Toruń (Poland).

E.P., J.Z., and M.M. acknowledge funding from the Research Council of Lithuania (LMTLT, grant No. S-LL-24-1).

The 1.2-m Kryoneri telescope is operated by the Institute for Astronomy, Astrophysics, Space Applications and Remote Sensing of the National Observatory of Athens

M.S. and G.D. acknowledge
support from the Astronomical station Vidojevica, funding
from the Ministry of Science, Technological Development and
Innovation of the Republic of Serbia (contract No. 451-03-66/
2024-03/200002) and by the EC through project BELISSIMA
(call FP7-REGPOT-2010-5, No. 256772).

S.A. is partially supported by the Fundamental Fund of Thailand Science Research and Innovation (TSRI) through the National Astronomical Research Institute of Thailand (Public Organization) (FFB680072/0269).

This work has made use of data from the European Space Agency (ESA) mission
{\it Gaia} (\url{https://www.cosmos.esa.int/gaia}), processed by the {\it Gaia}
Data Processing and Analysis Consortium (DPAC,
\url{https://www.cosmos.esa.int/web/gaia/dpac/consortium}). Funding for the DPAC has been provided by national institutions, in particular the institutions
participating in the {\it Gaia} Multilateral Agreement. 


\section*{Data Availability}

Our photometric data are available as supplementary material to this article. New BHTOM photometry is available from \url{http://bhtom.space/public/targets/TCrB}. 
Other photometric data used in this work are available directly from the AAVSO database and as supplementary material to \citet{2023MNRAS.524.3146S}.



\bibliographystyle{mnras}
\bibliography{example} 




\section*{Supporting information}
Supplementary data are available online.\\

\noindent \textbf{Table 3.} BHTOM photometry of T CrB in multiple filters.

\section*{Affiliations}
\noindent
{\it \small
$^{1}$Astronomical Institute of Charles University, V Holešovičkách 2, Prague, 18000, Czech Republic\\
$^{2}$Instituto de Astrofísica de Canarias, Calle Vía Láctea, s/n, E-38205 La Laguna, Tenerife, Spain\\
$^{3}$Astronomical Observatory, University of Warsaw, Al. Ujazdowskie 4, 00-478 Warsaw, Poland\\
$^{4}$Astrophysics Division, National Centre for Nuclear Research, Pasteura 7, 02-093 Warsaw, Poland\\
$^{5}$Departamento de Astrofísica, Universidad de La Laguna, E-38206 La Laguna, Tenerife, Spain\\
$^{6}$Astronomical Institute, University of Wrocław, ul. Mikołaja Kopernika 11, 51-622 Wrocław, Poland\\
$^{7}$Institute of Astronomy, Faculty of Physics, Astronomy and Informatics, Nicolaus Copernicus University in Toruń, Grudziądzka 5, 87-100 Toruń, Poland\\
$^{8}$Astronomical Observatory, Jagiellonian University, ul. Orla 171, 30-244 Kraków, Poland\\
$^{9}$Mt. Suhora Astronomical Observatory, University of the National Education Commission, ul. Podchorążych 2, 30-084 Kraków, Poland\\
$^{10}$Institute of Earth Systems, University of Malta, Msida MSD 2080, Malta\\
$^{11}$Znith Astronomy Observatory, Malta\\
$^{12}$Vereniging Voor Sterrenkunde (VVS), Zeeweg 96, B-8200 Brugge, Belgium\\
$^{13}$Bundesdeutsche Arbeitsgemeinschaft für Veränderliche Sterne, Munsterdamm 90, D-12169 Berlin, Germany\\
$^{14}$Flarestar Observatory, Fl.5 Ent.B, Silver Jubilee Apt, George Tayar Street, San Gwann, SGN 3160, Malta\\
$^{15}$Center for Astrophysics and Cosmology, University of Nova Gorica, Vipavska 11c, 5270 Ajdovščina, Slovenia\\
$^{16}$Astronomska revija Spika, Koprska ulica 94, 1000 Ljubljana, Slovenia\\
$^{17}$Institut de Ciències del Cosmos (ICCUB), Universitat de Barcelona (UB), Martí i Franquès 1, E-08028 Barcelona, Spain\\
$^{18}$Departament de Física Quàntica i Astrofísica (FQA), Universitat de Barcelona (UB), Martí i Franquès 1, E-08028 Barcelona, Spain\\
$^{19}$Institut d'Estudis Espacials de Catalunya (IEEC), Esteve Terradas, 1, Edifici RDIT, Campus PMT-UPC, 08860 Castelldefels, Barcelona, Spain\\
$^{20}$School of Physics, Trinity College Dublin, College Green, Dublin 2, Ireland\\
$^{21}$Horten Videregaende Skole, Strandpromenaden 33, 3183 Horten, Norway\\
$^{22}$Institute of Theoretical Physics and Astronomy, Vilnius University, Saulėtekio al. 3, Vilnius, 10257, Lithuania\\
$^{23}$Public Observatory Astrolab IRIS, Verbrandemolenstraat 5, 8901 Zillebeke, Belgium\\
$^{24}$Astronomical Observatory Institute, Faculty of Physics and Astronomy, Adam Mickiewicz University, ul. Słoneczna 36, 60-286 Poznań, Poland\\
$^{25}$IAASARS, National Observatory of Athens, Metaxa \& Pavlou St., GR-15236, Penteli, Athens, Greece\\
$^{26}$Astronomical Observatory, Volgina 7, 11060 Belgrade, Serbia\\
$^{27}$Faculty of Automatic Control, Electronics and Computer Science, Silesian University of Technology, Akademicka 16, 44-100 Gliwice, Poland\\
$^{28}$E. Kharadze Georgian National Astrophysical Observatory, 0301 Abastumani, Georgia\\
$^{29}$INAF – Osservatorio Astronomico di Padova, Vicolo dell'Osservatorio 5, I-35122 Padova, Italy\\
$^{30}$INAF – Osservatorio Astronomico di Brera, Via E. Bianchi 46, 23807 Merate (LC), Italy\\
$^{31}$National Astronomical Research Institute of Thailand (Public Organization), 260 Moo 4, Donkaew, Mae Rim, Chiang Mai 50180, Thailand\\
$^{32}$Sorbonne Université, CNRS, UMR 7095, Institut d'Astrophysique de Paris, 98bis Bd Arago, 75014 Paris, France\\
$^{33}$National and Kapodistrian University of Athens, Department of Physics, University Campus, Zografos GR-157 84, Athens, Greece
}





\bsp	
\label{lastpage}
\end{document}